# Résonance 2-surharmonique d'un contact unilatéral hertzien : approches théorique et expérimentale

**Joël Perret-Liaudet, Emmanuel Rigaud**

*Laboratoire de Tribologie et Dynamique des Systèmes, UMR 5513*
*Ecole Centrale de Lyon*
*69134 Ecully cedex*
*joel.perret-liaudet@ec-lyon.fr*

## Résumé :

*Pour un oscillateur à impact préchargé sous excitation surharmonique d'ordre 2, la non linéarité de Hertz constitue le précurseur à des réponses de type vibroimpact établies sur une large gamme de fréquence excitatrice. L'approche théorique retenue s'appuie sur une méthode balistique assortie d'une méthode de continuation type longueur d'arc. Les résultats expérimentaux obtenus à partir d'un banc expérimental dédié sont en accord avec les résultats théoriques. En particulier, le niveau seuil de l'excitation initiatrice de ces régimes dynamiques présentant des chocs est confirmé.*

## Abstract :

*For a preloaded impact oscillator under superharmonic excitation of order 2, the Hertzian non linearity constitutes the precursor of vibroimpacts established over a wide frequency range. The theoretical study is performed by using the shooting method with an arc length continuation. Experimental results obtained from a devoted test rig agree with theoretical ones. Particularly, the threeshold level of the excitation necessary to induce vibroimpacts is confirmed.*

## 1 Introduction

On considère le cas fondamental d'un contact sec de type hertzien, sans glissement, excité par une charge harmonique purement normale. Le montage expérimental mis en œuvre pour cette étude est constitué d'un double contact plan sphère plan permettant de décrire ce cas fondamental. Ce système est décrit Figure 1 et constitue un oscillateur à impact dont la raideur non linéaire de contact est déduite de la théorie de Hertz [1].

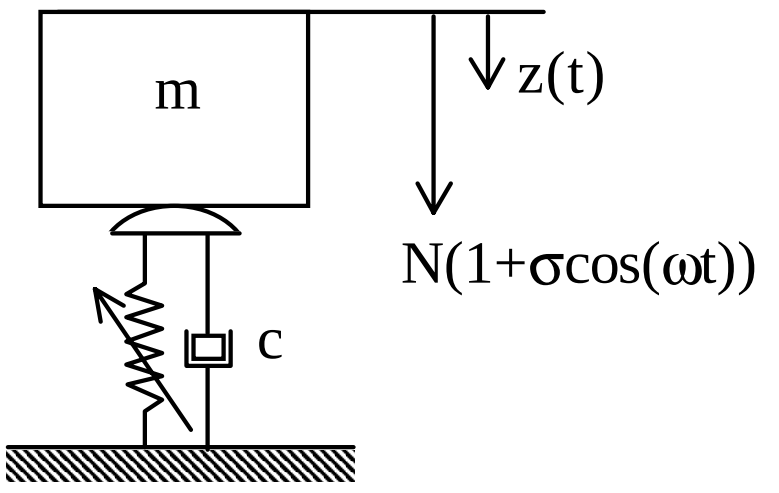

FIG.1 – Oscillateur non linéaire étudié

En introduisant un amortissement de type visqueux de viscosité *c*, l'équation régissant le mouvement de cet oscillateur s'écrit :

$$\begin{cases} m\ddot{z} + c\dot{z} + kz^{3/2} = N(1+\sigma\cos(\omega t)), & z \geq 0 \\ m\ddot{z} = N(1+\sigma\cos(\omega t)), & z < 0 \end{cases} \quad (1)$$

Dans cette équation, $z$ représente le déplacement de la masse $m$ ($z < 0$ pour un contact rompu) et $k$ est une constante prenant en compte la géométrie du contact et les caractéristiques des matériaux. $N$ constitue la charge statique exercée sur le contact. Les paramètres $\sigma$ et $\omega$ contrôlent l'amplitude et la pulsation de l'effort excitateur. La résonance principale d'un tel système a déjà été étudiée tant du point de vue théorique qu'expérimental [2-4]. Nous avons par ailleurs étudié les résonances de type sous- et surharmonique d'ordre 2 [5-6]. Les résultats théoriques obtenus montrent qu'elles constituent des précurseurs à l'apparition de régimes dynamiques présentant des pertes de contact. Dans ce contexte, l'objectif principal de ce travail a été de confirmer ces résultats par une campagne d'essais. Dans le cadre de cette présentation, seul le cas de la résonance surharmonique est reporté. Nous revenons dans un premier temps sur l'approche théorique, décrite au paragraphe 2, puis nous présentons l'approche expérimentale au paragraphe 3. Enfin, les résultats obtenus font l'objet du paragraphe 4.

## 2 Approche théorique

Introduisons les variables adimensionnelles suivantes :

$$\Omega^2 = \frac{3}{2}k^{2/3}N^{1/3}, \; q = \frac{3}{2}\left(z(k/m)^{2/3} - 1\right), \tau = \Omega\, t, \varpi = \frac{\omega}{\Omega}, \zeta = \frac{c\Omega}{2m} \quad (2)$$

On notera que $\Omega$ constitue la pulsation propre linéarisée de contact. L'équation (1) peut alors s'écrire sous la forme adimensionnelle suivante :

$$\begin{cases} \ddot{q} + 2\zeta\dot{q} + (1+2q/3)^{3/2} = 1+\sigma\cos(\varpi\tau), & q \geq -3/2 \\ \ddot{q} = 1+\sigma\cos(\varpi\tau), & q < -3/2 \end{cases} \quad (3)$$

La résonance 2-surharmonique correspond alors au cas où $\varpi = 1/2$, soit encore $\omega = \Omega/2$. Pour l'étudier, il est possible d'utiliser la méthode des échelles multiples, mais dans ce cas, seule la non linéarité de Hertz peut être traitée [5]. C'est pourquoi nous avons également utilisé une méthode balistique (shooting method) appliquée à des sections de Poincaré [6]. Elle est assortie d'une procédure de continuation de type longueur d'arc. Il est ainsi possible de suivre les réponses avec ou sans chocs, stables ou instables. Numériquement, les points d'équilibre dans la section de Poincaré sont déterminés en résolvant l'équation (3) par une méthode d'intégration temporelle de type différences centrées. Les matrices Jacobiennes nécessaires au schéma de prédiction correction sont aussi obtenues selon une procédure numérique. Seuls les résultats issus de la méthode balistique sont reportés dans cette étude.

## 3 Approche expérimentale

Le dispositif expérimental est présenté Figure 2. Il s'agit d'une bille préchargée entre un cylindre et un plan horizontal. Le cylindre est mobile dans le sens vertical. La bille de masse négligeable vis-à-vis de celle du cylindre ($m$ = 6,48 kg) est soumise au poids de ce dernier $N$ = 63,6 N. Le diamètre de la bille est de 25 mm. Les deux contacts sont de type sphère plan et répondent aux hypothèses de Hertz [1]. Les matériaux sont identiques en acier SAE 52100. La fréquence propre de contact mesurée (269,6 Hz) est en bon accord avec celle calculée (276,4 Hz) avec un écart relatif inférieur à 3%. Par ailleurs, le taux d'amortissement visqueux mesuré est environ égal à 0,6%. Un pot excitateur permet d'exciter le contact de manière purement normale. La force excitatrice et la force transmise au bâti sont mesurées à l'aide de capteurs piézoélectriques. Compte tenu de l'environnement des capteurs, la bande passante d'analyse est bien supérieure à la fréquence propre de contact (plus de 2 kHz). Les deux premiers harmoniques de la réponse sont analysés en détection par sensibilité de phase.

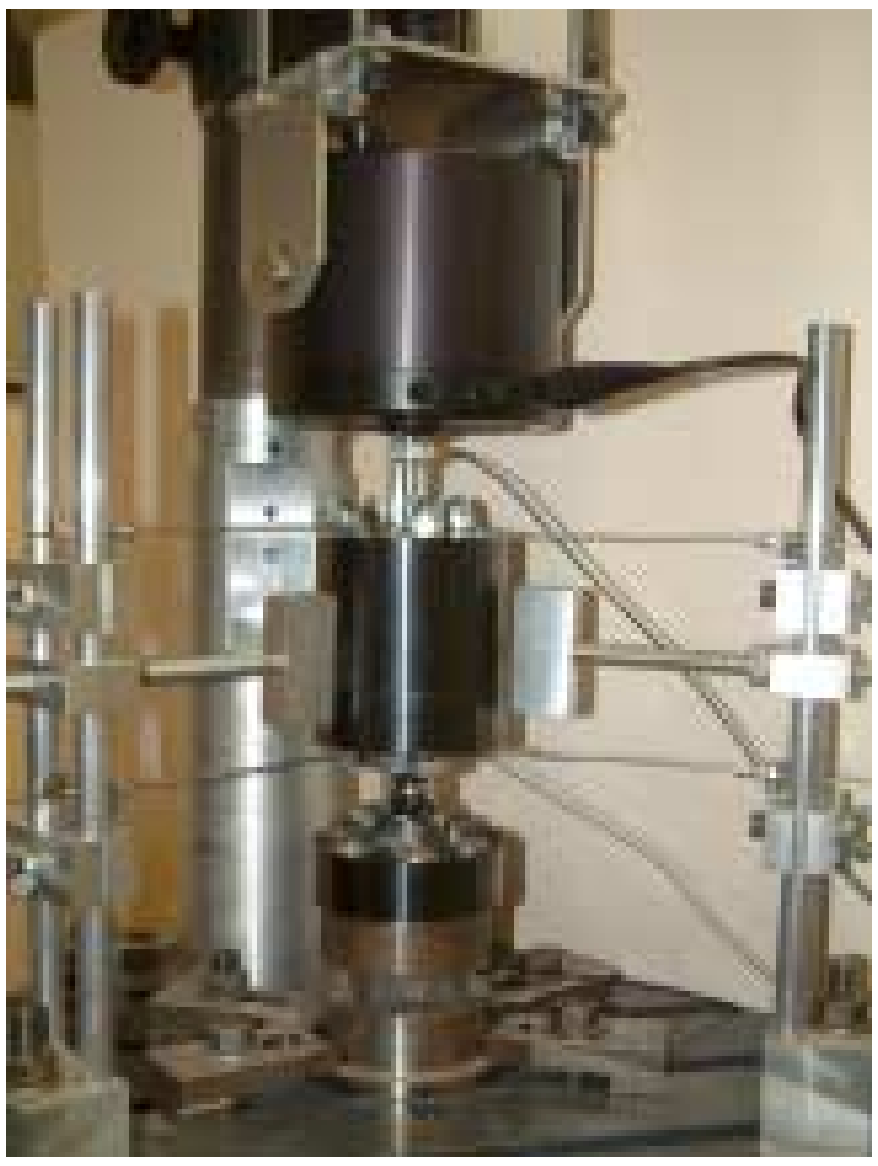

FIG.2 – Montage expérimental

# 4 Résultats

La figure 3 présente la résonance surharmonique en comparant les résultats théoriques et expérimentaux. Plus précisément, nous reportons les courbes de réponse en fréquence associées aux deux premiers harmoniques, de fréquences $\varpi$ et $2\varpi$, de l'effort transmis (amplitude et phase). Pour ce cas, seule la non linéarité de Hertz influe sur le comportement. En effet, les réponses dynamiques ne présentent aucune perte de contact. Le niveau de l'effort excitateur ($\sigma_{\text{expérimental}}$ = 28,5%) peut être considéré comme un niveau seuil car, au pic de résonance, l'amplification est telle que la réponse est proche de la perte de contact.

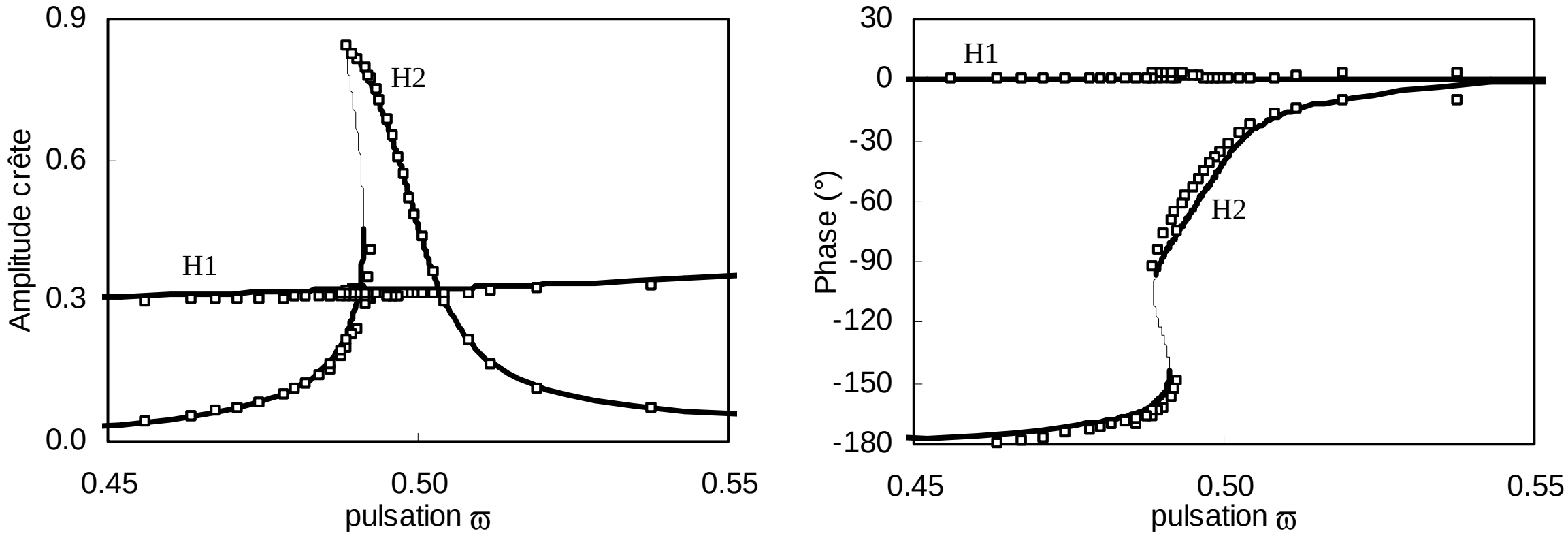

FIG.3 – Réponses en fréquence des deux premiers harmoniques H1, H2, de l'effort transmis (amplitude et phase). Trait épais : réponses théoriques stables ; trait fin : réponses théoriques instables ($\sigma_{\text{théorique}}$ = 24,5%) ; □ : résultats expérimentaux ($\sigma_{\text{expérimental}}$ = 28,5%).

En premier lieu, on peut noter une bonne adéquation entre résultats théoriques et expérimentaux, même si l'écart relatif portant sur le niveau excitateur est de l'ordre de 15%. Cet écart peut être dû à l'imprécision sur la valeur expérimentale de l'amortissement comme au fait que le niveau excitateur délivré par le pot excitateur n'est pas asservi. En second lieu, le comportement induit est de type mollissant en conformité avec celui observé pour la résonance principale. Par ailleurs, nous confirmons que c'est essentiellement l'harmonique 2 qui influe sur ce comportement. Soulignons enfin que, puisque les réponses sont sans choc, la résonance 2-surharmonique ne peut être prédite en approximant la raideur de contact au premier ordre, autour de la position d'équilibre statique.

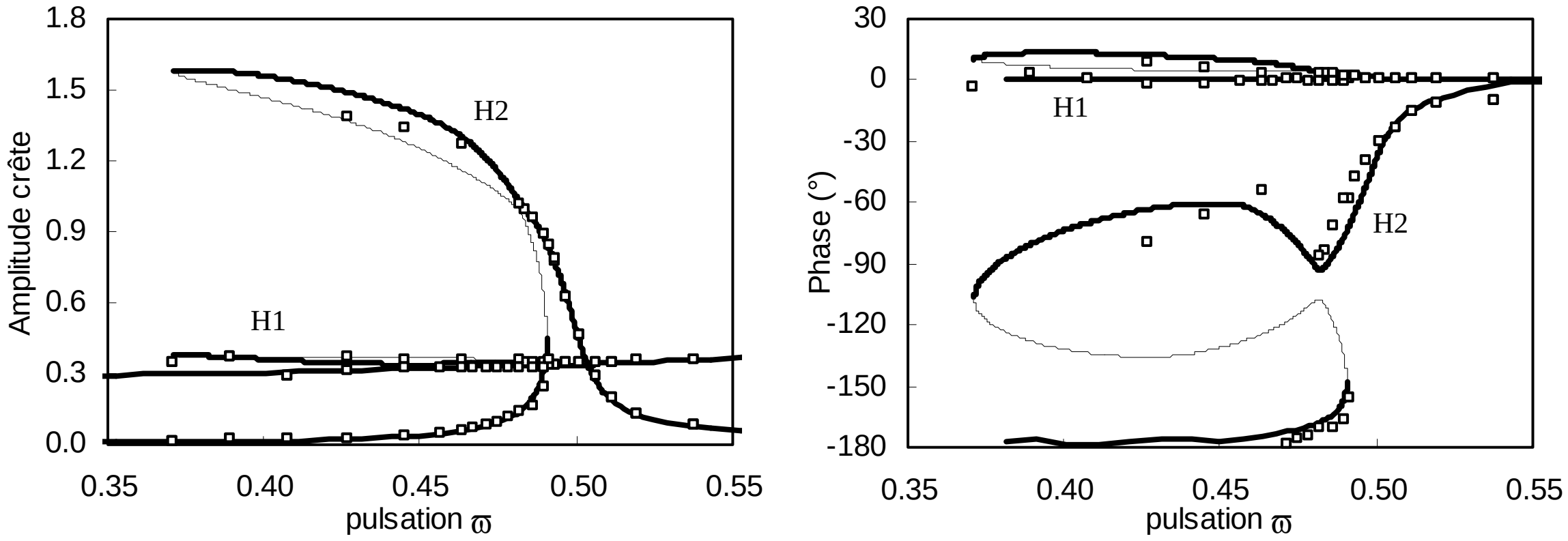

FIG.4 – Réponses en fréquence des deux premiers harmoniques H1, H2, de l'effort transmis (amplitude et phase). Trait épais : réponses théoriques stables ; trait fin : réponses théoriques instables ($\sigma_{\text{théorique}}$ = 26,5%) ; □ : résultats expérimentaux ($\sigma_{\text{expérimental}}$ = 29,5%).

Une légère augmentation du niveau exciteur (de 28,5% à 29,5% dans le cas expérimental) conduit à des réponses qui présentent maintenant des chocs. La Figure 4 illustre les résonances d'amplitude et de phase associées. Là encore, on note une bonne adéquation entre résultats théoriques et résultats expérimentaux. En premier lieu, la résonance conserve son caractère mollissant et s'établit sur une bande de fréquences plus large. Ainsi, la fréquence de saut théorique pour un balayage lent en descente est désormais de $\varpi$ = 0,372 au lieu de 0,488 relevée pour le cas de charge précédent. Du point de vue expérimental, elle passe de 131 à 115 Hz. Cet élargissement est soudain et peut s'expliquer par l'existence d'une bifurcation instable de type transcritique. Cette bifurcation correspond à la collusion entre le pic de résonance surharmonique induit par la non linéarité Hertzienne et décrit pour le cas de charge précédent, et un isola dont les réponses présentent essentiellement des chocs.

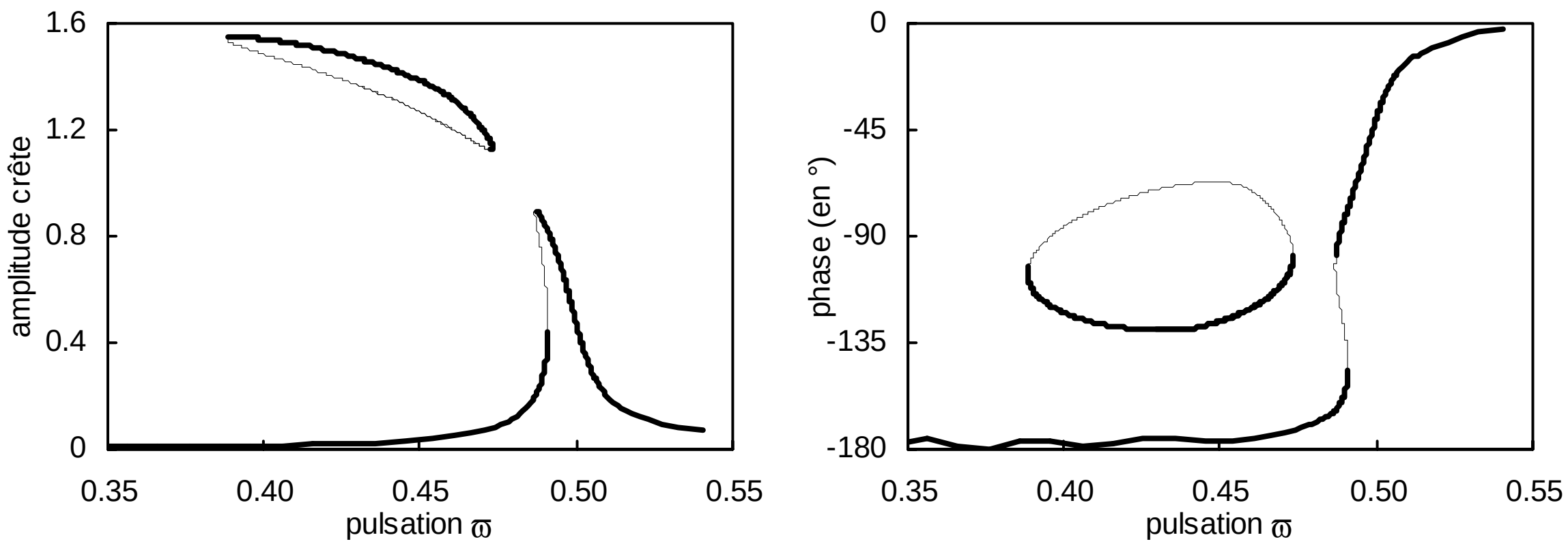

FIG.5 – Courbe numérique de la réponse en fréquence de l'harmonique 2 de l'effort transmis. Trait épais : réponses stables, trait fin : réponses instables ($\sigma_{\text{théorique}}$ = 25%).

En d’autres termes, les deux bifurcations rasantes associées à ces deux régimes dynamiques coalescent en une bifurcation transcritique. Ce résultat est bien illustré sur la Figure 5, où l’on observe la préexistence théorique de l’isola. La résonance de phase observée au cours des essais permet également de conclure sur ce scénario. En effet, on peut identifier clairement, Figure 4 , l’existence de cet isola.

Pour conclure, on peut donc dire que la non linéarité de Hertz constitue un précurseur à l’apparition de vibroimpacts. Là encore, il n’est pas possible de prédire ce phénomène si on linéarise la raideur de contact autour de la position d’équilibre statique. En effet, avec une raideur de choc linéaire, il existe bien un isola caractéristique de réponses avec chocs, mais la bifurcation transcritique ne peut être décrite.

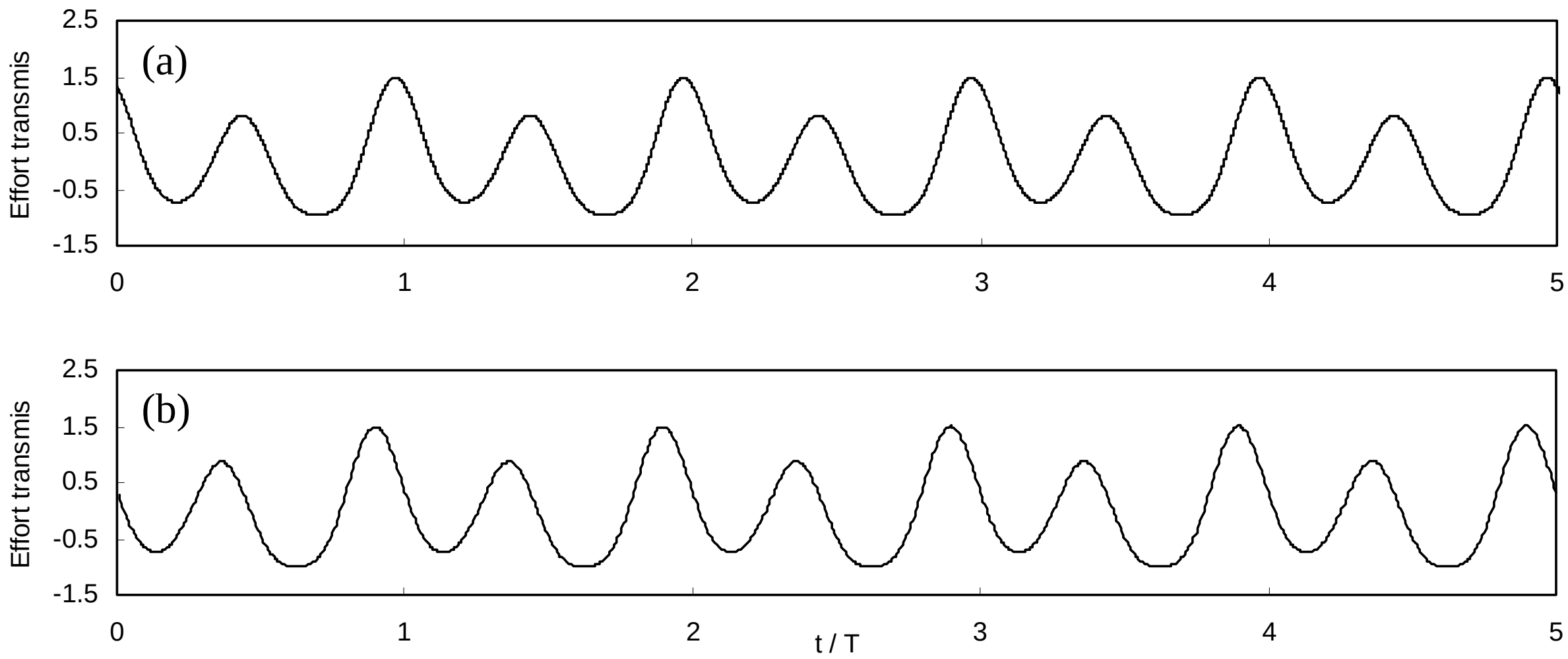

FIG.6 – Evolution temporelle de l’effort transmis observée pour $\varpi$ = 0,484. *T* constitue la période excitatrice. (a) : résultat expérimental ; (b) résultat théorique.

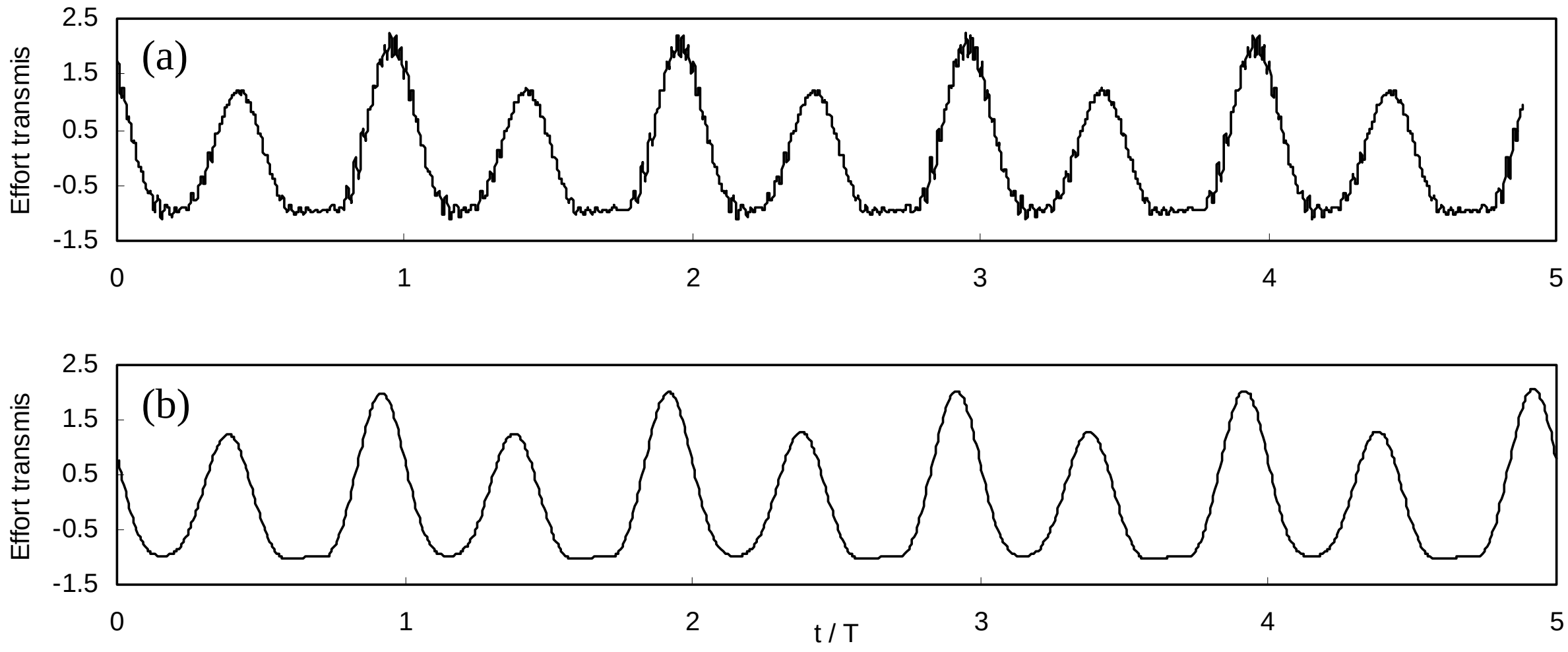

FIG.7 – Evolution temporelle de l’effort transmis observée pour $\varpi$ = 0,463. *T* constitue la période excitatrice. (a) : résultat expérimental ; (b) résultat théorique.

Enfin, les Figures 6 et 7 décrivent les évolutions temporelles de la réponse hors et avec chocs. Les résultats numériques sont en très bon accord avec les résultats expérimentaux, notamment sur l'alternance des pics. Les niveaux atteints sont importants. Dans le cas du régime de type vibroimpacts présenté Figure 7, le niveau crête de l'effort transmis est 2,5 fois plus élevé que l'effort statique. Les conséquences peuvent donc être préjudiciables en terme tribologique. Par exemple, ces surcharges peuvent conduire à des plastifications en sous couche au niveau du contact. Par ailleurs, la présence de chocs est susceptible d'exciter les modes supérieurs. C'est ce que l'on observe sur la courbe expérimentale de la Figure 7 (composantes haute fréquence). En conséquence, on constate un enrichissement significatif du bruit rayonné dans une gamme de fréquences où l'oreille est la plus sensible. Par exemple, sur notre banc d'essais, cette amplification est telle qu'elle permet de discerner à l'oreille et sans hésitation les régimes avec chocs.

## 5 Conclusion

Pour un contact sphère-plan, la méthode balistique assortie d'une technique de continuation permet de décrire la résonance 2-surharmonique. Les résultats théoriques sont en bon accord avec les résultats expérimentaux obtenus. La résonance induite par la seule non linéarité de Hertz est de type mollissant. Ce caractère est renforcé par la non linéarité de jeu. Au-delà d'un certain niveau excitateur, l'accrochage de la résonance conduit pour un balayage lent à des régimes de type vibroimpacts. Dans ce cas, l'élargissement de la résonance est soudain avec des niveaux de réponses élevés, conséquence d'une bifurcation de type transcritique. Le niveau seuil théorique au-delà duquel apparaissent ces réponses est en accord avec celui observé au cours des essais. Ce qu'il y a lieu de retenir, c'est que l'apparition de tels régimes ne peut être convenablement prédite, si l'on se contente de ne retenir que la raideur linéarisée de contact pour décrire la réponse. Plus généralement, au-delà du contact Hertzien, il apparaît nécessaire de prendre en compte de manière fidèle la non linéarité de contact, si l'on souhaite décrire convenablement la résonance surharmonique. Cette conclusion est aussi valide pour le cas de la résonance sousharmonique, dont la description fera l'objet d'un autre papier.

## Références